# Localized Magnetic States of Fe, Co, and Ni Impurities on Alkali Metal Films

P. Gambardella,[1] S. S. Dhesi,[2] S. Gardonio,[3] C. Grazioli,[3] P. Ohresser,[4] and C. Carbone[3]

[1]*Institut de Physique des Nanostructures, Ecole Polytechnique Fédérale de Lausanne, CH-1015 Lausanne, Switzerland*
[2]*European Synchrotron Radiation Facility, BP 220, F-38043 Grenoble, France*
[3]*Istituto di Struttura della Materia, Consiglio Nazionale delle Ricerche, Area Science Park, I-34012 Trieste, Italy*
[4]*LURE, Centre Universitaire Paris Sud, BP 34, F-91898 Orsay, France*

X-ray absorption spectroscopy (XAS) and x-ray magnetic circular dichroism (XMCD) have been used to study transition metal impurities on K and Na films. The multiplet structure of the XAS spectra indicates that Fe, Co, and Ni have localized atomic ground states with predominantly $d^7$, $d^8$, and $d^9$ character, respectively. XMCD shows that the localized impurity states possess large, atomiclike, magnetic orbital moments that are progressively quenched as clusters are formed. Ni impurities on Na films are found to be nonmagnetic, with a strongly increased $d^{10}$ character of the impurity state. The results show that the high magnetic moments of transition metals in alkali hosts originate from electron localization.

The ground state of an isolated transition metal atom possesses large spin and orbital magnetic moments given by Hund's rules. In the solid state, however, hybridization and crystal field effects strongly reduce both these magnetic moments. Dilute transition metal impurities in nonmagnetic hosts can be viewed as a bridge between the atomic and solid state and have consequently attracted a great deal of attention ever since the work of Friedel [1]. The extent to which the transition metal $d$ states interact with the valence bands of the host [1,2] directly influences macroscopic properties resulting in anomalies in the electronic transport, magnetic susceptibility, and specific heat. In this respect, alkali metals, with their simple electronic structure, are considered to be ideal hosts for studying the interaction between localized $3d$ states and a free-electron Fermi gas [3–11].

More than a decade ago, Riegel *et al.* and Kowallik *et al.* [3,4] showed that isolated Fe and Ni impurities implanted in K, Rb, and Cs hosts yield surprisingly large magnetic susceptibilities. Localized Fe $3d^6$ and Ni $3d^9$ configurations giving total magnetic moments of 6.7 and 3.5$\mu_B$, respectively, were indicated as the origin of these effects. Recently, however, anomalous Hall resistance measurements [5] of Fe and Co impurities on Cs thin films reported large magnetic moments which were interpreted in terms of polarized Cs conduction electrons, analogous to the mechanism giving rise to giant magnetic moments in CoPd and FePd alloys [12]. This interpretation was subsequently questioned [6] and not reproduced by first-principles band structure calculations [10,11]. The ground state electronic structure and magnetism of transition metal impurities in alkali hosts therefore remains an open question, in particular with respect to the degree of $3d$ localization, the local moment at the impurity sites, and the orbital contribution to the total magnetic moment. X-ray absorption spectroscopy (XAS) combined with x-ray magnetic circular dichroism (XMCD) provides an ideal approach to resolve these issues. The line shape of the XAS spectra is a fingerprint for the $d$-state configuration, whereas XMCD yields the spin and orbital moments via straight forward sum rules [13,14] in an element specific manner.

In this Letter, XAS and XMCD are used to probe the local electronic and magnetic structure of Fe, Co, and Ni impurities deposited on K and Na films. XAS convincingly shows that transition metals form highly localized atomic configurations in alkali metal hosts while sum rule analysis of the XMCD yields orbital magnetic moments comparable to the atomic limit. From the multiplet structure of the XAS at the Fe, Co, and Ni $L_{2,3}$ edges the atomic configurations are identified as $d^7$, $d^8$, and $d^9$, respectively. The quenching of the Ni magnetic moment in Na films is shown to be related to an increased $d^{10}$ weight with respect to the $d^9$ configuration. Moreover, the present measurements demonstrate that XMCD has a significant potential for the study of dilute systems with impurity concentration as low as $3 \times 10^{12}$ atoms cm$^{-2}$.

The experiments were performed at beam line ID8 of the European Synchrotron Radiation Facility in Grenoble. K and Na films were evaporated onto a clean Cu(111) substrate; transition metals were subsequently deposited in minute quantities, 0.002–0.015 monolayers (1 ML = $1.6 \times 10^{15}$ atoms cm$^{-2}$), at $T = 10$ K in order to obtain isolated impurities. The coverage of the transition metals was calibrated by measuring the in-plane remanent magnetization on Cu(111) at room temperature, which has a sharp onset at 2 ML [15]. The pressure measured during metal evaporation was $1.0 \times 10^{-10}$ mbar. Residual gas contamination was always lower than that detectable by O $K$-edge spectra recorded before and after metal evaporation. XAS at the $L_{2,3}$ edges was performed in total electron yield mode using circularly polarized (**P**) light with 99% polarization in magnetic fields up to $\pm 7$ T with the sample at $T = 10$ K. XMCD was recorded by switching both **P** and the sample magnetization (**M**).

Figure 1 shows the XAS spectra recorded for Fe, Co, and Ni impurities deposited on a K film for parallel (solid lines) and antiparallel (dashed lines) alignment of **P** with **M**. For clarity, these spectra are referred to as $I_+^{exp}$ and $I_-^{exp}$, respectively. The pairs of XAS spectra were normalized to the incident photon flux and to each other at the preedge. A common linear baseline was subtracted from each set of spectra after normalization. The XMCD spectra, given by $I_+^{exp} - I_-^{exp}$, are also shown for each case. For each transition metal there are several notable features. The XAS spectra present narrow multiplet structures which are not observed in the corresponding bulk metal spectra [16,17]. This is a clear indication of 3d localization on the transition metal impurities. In addition, the magnitude of the XMCD is significantly larger compared to the bulk [16,17] and compared to that reported for low-dimensional structures, where the size of the magnetic moments is increased due to the reduced coordination [18]. Further, in the case of Fe and Co, the XMCD at the $L_2$ edge has the opposite sign with respect to bulk spectra, whereas it is zero for Ni.

To determine the electronic configurations of the localized states, the multiplet structure of the XAS and XMCD spectra is compared to atomic multiplet calculations reported by van der Laan and Thole [19]. The insets of Figs. 1(a) and 1(b) show the XAS and XMCD for the $3d^n \rightarrow 2p^5 3d^{n+1}$ absorption thresholds calculated for $d^7$ and $d^8$ configurations, respectively, in zero crystal field with an atomic value for the spin-orbit splitting. The calculated spectra are labeled $I_+^{th}$ and $I_-^{th}$, for comparison with the corresponding experimental spectra. Clearly there is a very close match between the theoretical and experimental XAS spectra of Figs. 1(a) and 1(b). In the case of $I_-^{exp}$ (Fe), five peaks can be distinguished and one additional peak appears for $I_+^{exp}$ (Fe). All the structure in the XAS spectra of Fig. 1(a) is reproduced in the calculated spectra for a $d^7$ ion shown in the inset of Fig. 1(a). However, one significant difference is that all the higher energy structure of $I_-^{th}$ (Fe) is absent for $I_+^{th}$ (Fe). The fact that this structure appears in $I_+^{exp}$ (Fe) is mainly due to incomplete alignment of the Fe magnetization with the applied field. Magnetization loops (not shown) indicate that **M** is not saturated at 7 T. Other minor differences between the experimental and calculated line shapes might be ascribed to a small fraction of the spectral weight arising from a mixing of different initial state configurations. Similar arguments can be used to understand the $d^8$ ground state for the Co case shown in Fig. 1(b). The line shape of $I_-^{exp}$ (Co) corresponds to the calculated [19] multiplet structures of a $d^8$ ground state shown as $I_-^{th}$ (Co) in the inset of Fig. 1(b). For a $d^8$ ground state $I_+^{th}$ is zero in the atomic limit; accordingly, we find that all peaks have a strongly reduced intensity in $I_+^{exp}$ (Co) compared to $I_-^{exp}$ (Co). The Ni case is very simple since the single peak at the $L_3$ edge indicates that the only states which are excited in the absorption process are those arising from the $j = 3/2$ initial state. Dipole selection rules, in this case, imply that the empty d states are of purely $j = 5/2$ character indicating a $d^9$ atomic ground state with $S = 1/2$, $L = 2$, and $J = 5/2$ [20].

In the present study, XAS combined with XMCD clearly demonstrates a predominant $d^7$ configuration for Fe adatoms on K. Previous studies [3,8] were unable to discriminate between a Fe $d^6$ or $d^7$ configuration since the size of the effective moment is the same for each configuration. One could argue that differences in the impurity coordination might strongly affect the ground state configurations of Fe, since we have investigated Fe impurities on K films rather than *in* the films. These effects, however, would rather favor an additional charge transfer from the electropositive host to implanted impurities, thus leading to an increase of the $d$ electron count. For Co, a $d^7$ configuration has been suggested [6] in order to interpret the anomalous Hall resistance measurements [5] on alkali films. However, here we present strong evidence for a Co $d^8$ ground state, in very good agreement with recent Hubbard-modified local spin density calculations for Co impurities in Cs films [10]. In the case of Ni, the $d^9$ configuration agrees with the interpretation of perturbed $\gamma$-ray distribution data for Ni impurities in K, Rb, and Cs hosts [4]. The atomic configurations, the ground state $S$, $L$, $J$ values, and the predicted atomic magnetic moments for Fe, Co, and Ni are summarized in Table I.

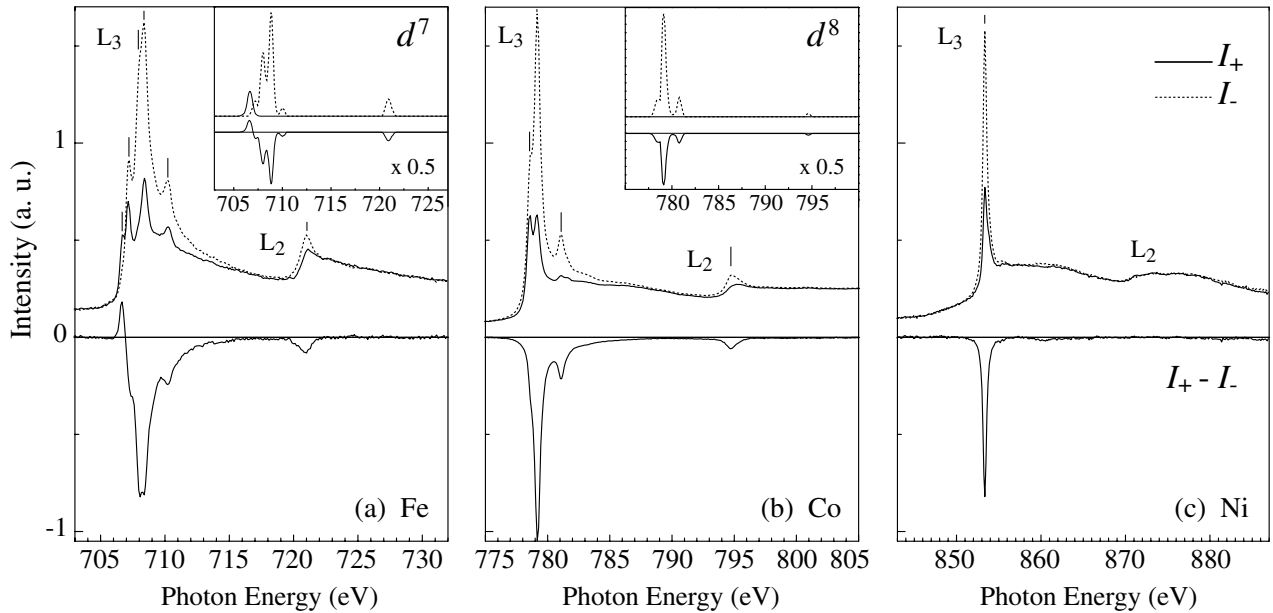

FIG. 1. XAS spectra over the $L_{2,3}$ edges recorded with **P** parallel (solid line) and antiparallel (dashed line) to **M** for (a) 0.015 ML Fe, (b) 0.015 ML Co, and (c) 0.004 ML Ni deposited on K films. The insets of (a) and (b) show the corresponding spectra calculated [19] for $d^7$ and $d^8$ atomic configurations, respectively (the energy scale has been renormalized to match the experimental $L_3$-$L_2$ separation). In each case the resulting XMCD ($I_+ - I_-$) is also shown.

TABLE I. Calculated Hund's rules ground state values of the isotropic spin moment ($S$), dipole spin moment ($T$), orbital moment ($L$), total angular moment ($J$), and magnetic moment $m = g_J \mu_B \sqrt{J(J+1)}$ corresponding to the $3d$ configurations identified on K films. The theoretical and experimental $R$ values are also compared for 0.015 ML Fe, 0.015 ML Co, and 0.004 ML Ni.

| State | $S$ | $T$ | $L$ | $J$ | $m$ | $R_{th}$ | $R_{exp}$ |
|---|---|---|---|---|---|---|---|
| Fe $d^7$ | 3/2 | $-1/7$ | 3 | 9/2 | $6.63\mu_B$ | 3/2 | $0.95 \pm 0.05$ |
| Co $d^8$ | 1 | 1/7 | 3 | 4 | $5.59\mu_B$ | 1 | $0.89 \pm 0.04$ |
| Ni $d^9$ | 1/2 | 2/7 | 2 | 5/2 | $3.55\mu_B$ | 2/3 | $0.67 \pm 0.05$ |

The expectation values of the orbital moment $\langle L_z \rangle$, spin moment $\langle S_z \rangle$, and magnetic dipole term $\langle T_z \rangle$ are related by the sum rules [13,14] to the integrated XMCD signal. Here we use the ratio $R$, which is given by

$$R \equiv \frac{\langle L_z \rangle}{2\langle S_z \rangle + 7\langle T_z \rangle} = \frac{2}{3} \frac{\Delta A_3 + \Delta A_2}{\Delta A_3 - 2\Delta A_2}, \quad (1)$$

where $\Delta A_2$ and $\Delta A_3$ are the integrated XMCD signals over the $L_{2,3}$ edges, respectively. Table I compares the theoretical ($R_{th}$) and experimental ($R_{exp}$) values of $R$. We find a very good agreement between $R_{exp}$ deduced from the XMCD measurements and $R_{th}$ calculated in the atomic limit for Co and Ni. For Fe there is a discrepancy which could be due to finite temperature effects [20]; we note that $R \approx 1$ is also found for the calculated $d^7$ spectra [19]. The application of the XMCD sum rules in the present context is particularly relevant, since their derivation stems from a localized atomic model [13,14]. So far, their validity has only been checked for bulk magnetic systems [16]. In particular, for atomic Fe and Ni, we note the importance of $\langle T_z \rangle$ which is generally assumed to be negligible in bulk $3d$ systems [16].

Figure 2 shows $R_{exp}$ determined from the XMCD spectra as a function of increasing Co and Ni impurity coverage on K and Na films. For Co/K, Co/Na, and Ni/K, $R_{exp}$ begins to decrease above 0.015 ML due to the formation of clusters by statistical deposition on neighboring adsorption sites. Monte Carlo simulations on a periodic lattice show that at 0.005 (0.015) ML the mean cluster size is 1.06 (1.20) atoms, whereas at 0.15 ML it increases to 8.2 atoms. With growing cluster size, $d$-state hybridization quenches the orbital moment. In parallel, the multiplet structure in the XAS spectra is broadened and eventually resembles that of the bulk metal. The Co XMCD shown in Fig. 3 clearly reflects this trend. The $L_3$ peak at 779 eV broadens, whereas the smaller satellite at 781 eV becomes less defined and finally disappears. At the $L_2$ edge, the effects of hybridization are even more spectacular. The XMCD is negative first, as predicted by an atomic model [19], and gradually becomes increasingly positive until it resembles the XMCD for bulk Co.

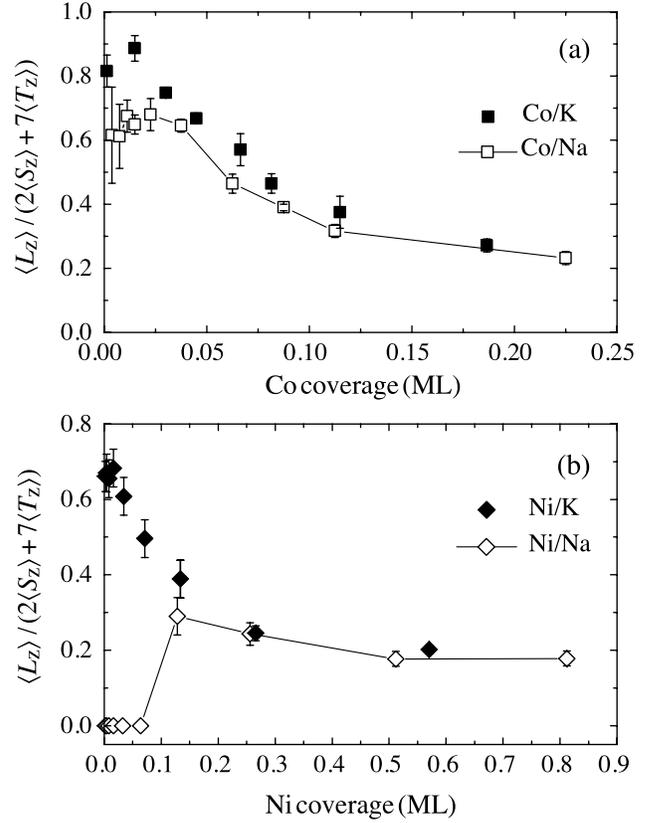

FIG. 2. $R_{exp}$ as a function of coverage determined for (a) Co on K and Na and for (b) Ni on K and Na.

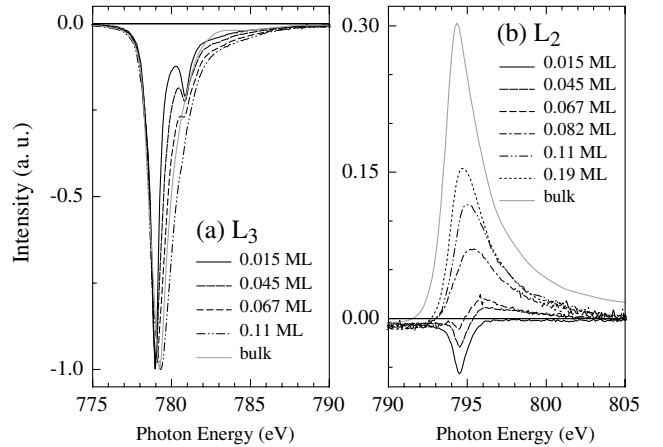

FIG. 3. XMCD spectra recorded for Co on K showing the line shape changes as the atomic Co forms clusters with increased coverage.

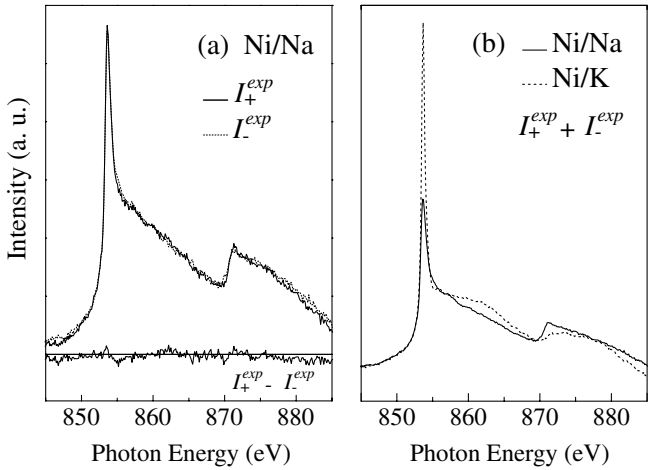

FIG. 4. (a) XAS spectra recorded with **P** parallel (solid line) and antiparallel (dashed line) to **M** for 0.004 ML Ni on Na and related XMCD. (b) Comparison between the $I_+^{exp} + I_-^{exp}$ spectra, normalized to the $s$ edge jump, for 0.004 ML Ni on Na (solid line) and on K (dashed line).

Previous investigations [3,4,7] reported deviations from a purely atomic magnetic behavior as the host moves up from K to Li in the alkali series in the periodic table. The observed reduction of the Co/Na $R_{exp}$ shown in Fig. 2 compared to Co/K indicates a reduction of the orbital magnetic moment due to the increased hybridization of the $3d$ states with the host. More striking, Ni impurities on Na do not present any XMCD up to a coverage of $\sim 0.012$ ML implying a nonmagnetic ground state. This effect has been explained [9] in terms of a broadening of the impurity state associated with the increased electron density of Na hosts compared to heavier alkali elements. Alternative explanations such as Kondo-type demagnetization have been excluded in earlier work [4]. Here we argue that the Ni impurity states in Na assume a significant $d^{10}$ character, and that this may have a direct bearing on the quenching of the Ni moment. Figure 4(b) compares the isotropic $(I_+^{exp} + I_-^{exp})$ XAS spectra for 0.004 ML of Ni on Na (solid line) and on K (dotted line). The weak $L_2$ intensity in the Ni/Na spectra indicates the presence of a $j = 3/2$ component in the Ni final states likely due to moderate hybridization with the host. The intensity of the $L_3$ edge due to the $3d^9 \rightarrow 2p^53d^{10}$ transition is reduced by more than 50% for Ni/Na compared to Ni/K. The width of the line, however, although larger compared to Ni/K (0.80 vs 0.52 eV FWHM) is still much narrower relative to bulk Ni spectra ($\sim 1.7$ eV [17]), indicating that Ni valence states can still be considered as predominantly localized. These observations suggest the presence of two resonant $d^9$ and $d^{10}$ configurations of ionic type [21], where fast incoherent electron hopping (i.e., charge fluctuation) forbids the existence of a magnetic moment.

In conclusion, $3d$ transition metal impurities on alkali films display localized atomic configurations with fully unquenched orbital magnetic moments. The present results conclusively demonstrate that the high moments in alkali systems originate from the localization of the $3d$ states.

We thank K. Larsson for his skilled technical assistance.